\documentclass[twocolumn,showpacs,preprintnumbers,amsmath,amssymb]{revtex4}

\usepackage{graphicx}
\usepackage{dcolumn}
\usepackage{bm}

\def\beq{\begin{equation}}
\def\eeq{\end{equation}}
\def\bea{\begin{eqnarray}}
\def\eea{\end{eqnarray}}

\def\nn{\nonumber}
\def\Eq#1{Eq.~(\ref{#1})}

\begin{document}

\preprint{IFIC/09-25}

\title{Constraining heavy colored resonances from top-antitop quark events}

\author{Paola Ferrario}\email{paola.ferrario@ific.uv.es}
\author{Germ\'an Rodrigo}\email{german.rodrigo@ific.uv.es}
\affiliation{Instituto de F\'{\i}sica Corpuscular, 
Consejo Superior de Investigaciones Cient\'ificas -- UVEG, 
Apartado de Correos 22085, E-46071 Valencia, Spain.}


\date{June 30, 2009}

\begin{abstract}
Recent measurements of the top quark charge asymmetry at Tevatron 
disfavor the existence of flavor universal axigluons and colorons 
at $2\sigma$. In this letter we explore the possibility of 
reconciling the data with these models and use the charge asymmetry 
and the invariant mass distribution of top-antitop quark pair events
to constrain the mass and couplings of massive color-octet gauge 
bosons decaying to top quarks. 
\end{abstract}

\pacs{14.65.Ha, 11.30.Er, 12.10.Dm}

\maketitle

The top quark, being the heaviest known elementary particle, plays 
a fundamental role in many extensions of the Standard Model (SM)
and in alternative mechanisms for electroweak symmetry breaking (EWSB). 
Since its discovery in 1995 at Tevatron, many properties of the top
quark, such as mass and total cross section, have been measured with 
high precision~\cite{:2009ec}, 
also allowing for limits to be set on physics beyond the SM. 
Because the production cross section of top quarks is about 50 to 100 
times larger at LHC than at Tevatron, the LHC will produce, even with 
early data, more top-antitop quark pairs than the Tevatron during 
its whole life
-- $200$~pb$^{-1}$ of integrated luminosity at $10$~TeV center of mass 
energy are expected to be collected by the end of 2010 --  
offering new opportunities to probe new physics in the top quark 
sector. 

In this paper, we shall set bounds on the mass and couplings of 
heavy colored gauge bosons decaying to top quarks by analyzing recent 
measurements of the top-antitop quark pair invariant 
mass distribution~\cite{Aaltonen:2009iz} and the charge asymmetry 
(or forward-backward asymmetry)~\cite{cdf,newcdf,d0}. 
Particularly interesting is 
the fact that the uncertainty of both measurements is still statistically 
dominated, which opens the possibility for further improvements in 
the near future even before the start of the LHC. 

Several models predict the existence of new electroweak $W'$ and $Z'$
gauge bosons, color-octet gauge bosons, or gravitons that should 
be detectable in top-antitop quark events, particularly in those 
models where the coupling of the new gauge bosons to the third 
generation is enhanced with respect to the lighter fermions. 
The most stringent lower bounds on the mass of such new states
are about $800$~GeV for the $W'$ and 
$Z'$~\cite{Aaltonen:2008dn,Aaltonen:2009qu,:2007dia,d0ttbar}, 
$1.2$~TeV for axigluons and flavor-universal colorons~\cite{Aaltonen:2008dn}, 
and $600$~GeV for gravitons~\cite{graviton}. 
Electroweak precision measurements rise the exclusion mass 
region of the $Z'$ to above $3$~TeV in Randall-Sundrum 
scenarios~\cite{Agashe:2003zs}.
Those limits, however, should be taken with care as 
they depend on the given model adopted to set that bounds, 
although the numbers quoted above are quite similar across
different analysis. 

We are interested in color-octet gauge bosons which couple to 
quarks with a nonvanishing axial-vector coupling. 
Those states appear, for example, in 
chiral color models~\cite{chiralcolor} 
where the SM color group 
have been extended to $SU(3)_R \otimes SU(3)_L$, and the symmetry 
breaking to the diagonal $SU(3)_C$ generates the massive axigluon, 
which couples to quarks with a pure axial-vector structure and 
the same strength as QCD. 
Chiral color models also require the existence of extra fermions 
to cancel anomalies, and extra Higgs bosons to break the enlarged 
gauge symmetry. We will assume that it is always possible to set them 
arbitrarily heavy. Those models can also be generalized
by considering different coupling constants associated with each 
$SU(3)$ component~\cite{Cuypers:1990hb,Carone:2008rx}, 
thus generating both vector and axial-vector couplings 
of the axigluon to quarks. 

We shall not stick here to a particular model, but will analyze the most 
general scenario where the heavy resonance interacts with quarks 
with arbitrary vector $g_V$ and axial-vector $g_A$ strength  
relative to the strong coupling $g_S$. We also assume that 
there is no direct coupling of a single resonance to an even number 
of gluons, and therefore the production of top quarks is driven 
by $q\bar q$ events. 
This choice is motivated by different implementations of models 
predicting the existence of extra color-octet gauge bosons. For example, 
the asymmetric chiral color model~\cite{Cuypers:1990hb}
allows the existence of three axigluon vertices, which are forbidden 
in the usual chiral color model by parity~\cite{chiralcolor}, 
but exclude gluon-gluon-axigluon vertices as well. 
Models in extra warped dimensions, where Kaluza-Klein (KK) modes
can be single produced, have been constructed~\cite{Agashe:2007jb}, 
but in the conventional and more extended extra dimensional 
models, a single KK gauge field does not couple to two SM gauge bosons 
at leading order by orthonormality of field profiles~\cite{Dicus:2000hm}. 

The Born cross-section for $q\bar{q}$ annihilation into top quarks 
in the presence of a color-octet vector resonance reads
\bea
&& \frac{d\sigma^{q\bar{q}\rightarrow t \bar{t}}}{d\cos \hat{\theta}} =
\alpha_S^2 \: \frac{T_F C_F}{N_C} \:
\frac{\pi \beta}{2 \hat{s}}
\left\{ 1+c^2+4m^2  \right. \nn \\ && + \frac{2 \hat{s} (\hat{s}-m_G^2)}
{(\hat{s}-m_G^2)^2+m_G^2 \Gamma_G^2}
\left[ g_V^q \, g_V^t \, (1+c^2+4m^2) \right. \nn \\ && \left. 
+ 2 \, g_A^q \, g_A^t \, c  \right] +
\frac{\hat{s}^2} {(\hat{s}-m_G^2)^2+m_G^2 \Gamma_G^2}
\left[ \left( (g_V^q)^2+(g_A^q)^2 \right) \right. \nn \\ && \times 
\left( (g_V^t)^2 (1+c^2+4m^2) +  (g_A^t)^2 (1+c^2-4m^2) \right)
\nn \\ && \left.\left.
+ 8 \, g_V^q \, g_A^q \, g_V^t \, g_A^t \, c \, \right]
\right\}~,
\label{eq:bornqq}
\eea
where $\hat{\theta}$ is the polar angle of the top quark with respect
to the incoming quark in the center of mass rest frame,
$\hat{s}$ is the squared partonic invariant mass,
$T_F=1/2$, $N_C=3$ and $C_F=4/3$ are color factors,
$\beta = \sqrt{1-4m^2}$ is the velocity of the top quark,
with $m=m_t/\sqrt{\hat{s}}$, and $c = \beta \cos \hat{\theta}$.
The parameters $g_V^q (g_V^t)$ and $g_A^q(g_A^t)$ represent, 
respectively, the vector and axial-vector couplings of the
excited gluons to the light quarks (top quarks).
Color-octet resonances are naturally broad: 
\bea
\frac{\Gamma_G}{m_G} \approx \frac{\alpha_{S}\, T_F}{3}
\sum_{i=q,t} \left( (g_V^i)^2+(g_A^i)^2 \right)
\approx {\cal O} (10\%)~. 
\eea

The terms in \Eq{eq:bornqq} that are odd in $c$ generate a 
charge asymmetry, namely a difference in the differential 
distribution of top versus antitop quarks. 
At Tevatron, this charge asymmetry is equivalent to a 
forward-backward asymmetry
as a consequence of charge conjugation symmetry.
CP violation arising from electric or chromoelectric 
dipole moments of the top quark do not contribute to the
asymmetry, unless the asymmetry is defined through the 
decay products.  
Only the terms in \Eq{eq:bornqq} that are even in $c$ contribute 
to the top-antitop quark invariant mass distribution.  

At leading order in QCD, there is no charge asymmetry;
the differential distributions of top and 
antitop quarks are identical. But due to higher order radiative corrections
a charge asymmetry is generated 
at ${\cal O} (\alpha_S^3)$ in $q\bar q$ events, and top quarks
become more abundant in the direction of the incoming light quarks. 
The QCD prediction for Tevatron is~\cite{Antunano:2007da,Bowen:2005ap,mynlo}
\beq
A^{p\bar p} = \frac{N_t(y\ge 0)-N_{\bar t}(y\ge 0)}
{N_t(y\ge 0)+N_{\bar t}(y\ge 0)} = 0.051(6)~,
\label{appbar}
\eeq
where $y$ denotes the rapidity. 
This also includes a small mixed QCD-electroweak contribution. 
The charge asymmetry can also be defined through 
$\Delta y=y_t-y_{\bar t}$, which is equivalent to evaluate 
the asymmetry in the $t\bar t$ rest frame because $\Delta y$
is invariant under boosts. In that frame the asymmetry 
is about $50$\% larger~\cite{Antunano:2007da}: 
$A^{t\bar t} = 0.078(9)$.
Although one can enlarge the uncertainty of the QCD asymmetry 
to a conservative $30$\% in order to account for higher order corrections,
the result in \Eq{appbar} has been proven to be 
stable to threshold resummations~\cite{Almeida:2008ug}, 
which shift the central value only by one per mille. 
Whether one prefers to quote a more conservative 
theoretical prediction for the asymmetry or not, is not relevant 
at the moment, as the present uncertainty of the experimental 
measurement (see below) is of the same order as the size of the 
QCD prediction. In the following, we will use therefore  
the result in \Eq{appbar} as reference number.

\begin{figure}[t]
\includegraphics[width=8cm]{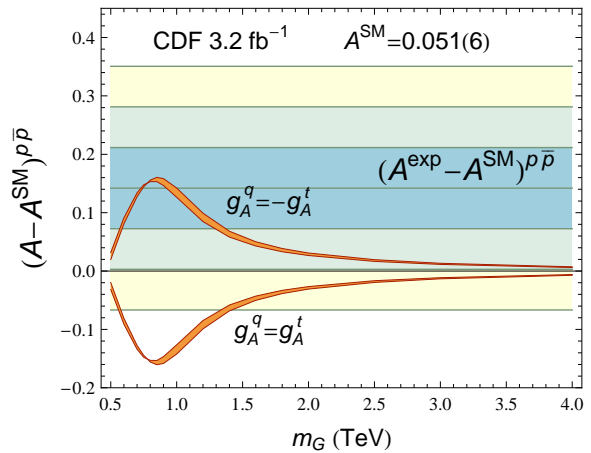}
\caption{\label{fig:anti} Comparison of the axigluon contribution to the 
top quark charge asymmetry with the $1\sigma$, $2\sigma$, and $3\sigma$
contours as a function of the axigluon mass. We also consider 
the case $g_A^q=-g_A^t=1$.}
\end{figure}

At Tevatron, CDF and D0 have recently measured the charge 
(or forward-backward) asymmetry  
with top-antitop quark events~\cite{cdf,newcdf,d0}. 
The measurement has been performed 
both in the $p\bar p$ rest frame~\cite{cdf,newcdf,d0} and in 
the $t\bar t$ rest frame~\cite{cdf}. 
The most recent measurement in the laboratory frame
with $3.2$~fb$^{-1}$ is~\cite{newcdf}
\beq
A^{p\bar p} = 0.193 \pm 0.065_{\, \rm stat.} 
\pm 0.024_{\, \rm syst.}~,
\label{eq:newcdf}
\eeq
to be compared with the one year old result:
$A^{p\bar p} = 0.17 \pm 0.07_{\, \rm stat.} \pm 0.04_{\, \rm syst.}$, 
with $1.9$~fb$^{-1}$~\cite{cdf}.
The uncertainty of both measurements is still large, but 
systematic errors have been improved considerably from 
one measurement to another, 
and statistical errors have decreased accordingly. 
Moreover, it turns out to be quite interesting that the 
uncertainty is still statistically dominated, 
and hence significant improvements should be expected in the 
near future. Indeed, we shall see that the new measurement 
have a larger impact in constraining heavy resonances than the 
older one. Comparing \Eq{appbar} with \Eq{eq:newcdf}, we can 
deduce that heavy resonances giving rise to a 
vanishing or negative charge asymmetry are disfavored 
at $2\sigma$ (see Fig.~\ref{fig:anti}). This is the case of 
colorons ($g_A=0$) and normal axigluons ($g_V=0,g_A=1$). 
At $3\sigma$ one can also exclude axigluon masses 
below $1.4$~TeV. In comparison with 2008, 
where at $2\sigma$ there was still a sizable room for a negative 
contribution to the asymmetry~\cite{Rodrigo:2008qe}, the situation 
has changed dramatically. 

Now We explore whether it is still possible to reconcile the axigluon 
with the measurement of the charge asymmetry. A positive asymmetry 
can be generated if the term from the squared amplitude of 
the massive color-octet in \Eq{eq:bornqq}, which is proportional to 
$8 g_V^q g_A^q g_V^t g_A^t c$, dominates over the term of 
the interference, that is proportional to $2 g_A^q g_A^t c$. 
This is possible if the vector couplings are large 
enough~\cite{Ferrario:2008wm}. However, although the total 
cross section might still be compatible with the SM prediction
in that case, because the contribution of the excited gluon 
is suppressed by powers of its mass, the top-antitop quark invariant 
mass distribution might be enhanced considerably, due to the factor 
\beq
\left( (g_V^q)^2 + (g_A^q)^2 \right)\left( (g_V^t)^2 + (g_A^t)^2 \right)~,
\eeq 
particularly for high values of the top-antitop quark invariant mass. 
The top-antitop quark invariant mass distribution 
has been measured very recently~\cite{Aaltonen:2009iz}.
The last bin 
\bea
\frac{d\sigma}{dM_{t\bar t}} (0.8-1.4~{\rm TeV}) &=&
0.068 \pm 0.032_{\, \rm stat.} \pm 0.015_{\, \rm syst.} \nn \\ &&  
\pm 0.004_{\, \rm lumi.}~
({\rm fb~GeV}^{-1})
\label{mttbar}
\eea
is, for the reasons explained above, the most sensible to 
extra contributions beyond the SM at the TeV scale. 

As in Ref.~\cite{Ferrario:2008wm}, we consider the flavor-universal 
scenario where light and top quarks share the same vector $g_V$ 
and axial-vector $g_A$ coupling to the massive color-octet gauge boson
and evaluate the size of the charge asymmetry for different 
values of the couplings and the mass. Then, we set the limits 
on the parameter space that are compatible with the newest 
experimental value (\Eq{eq:newcdf}), after subtracting the 
theoretical QCD prediction (\Eq{appbar}). 
We do the same exercise with the top-antitop quark invariant mass
distribution in the interval $800$~GeV$<M_{t\bar t}<1.4$~TeV (\Eq{mttbar}).
Within $1\sigma$ we allow the invariant mass distribution in 
that bin to be enhanced by $50$\%. The charge asymmetry and 
the invariant mass distribution probe different combinations
of the vector and axial-vector couplings; therefore by
combining both limits (we do not perform a global fit), 
one can constrain complementary regions of the parameter space. 
Similar analyses have also been performed recently in 
warped extra dimensional models~\cite{Djouadi:2009nb}
and in the asymmetric chiral color model~\cite{Martynov:2009en}. 

\begin{figure}[t]
\vspace{-.5cm}
\includegraphics[width=6.7cm]{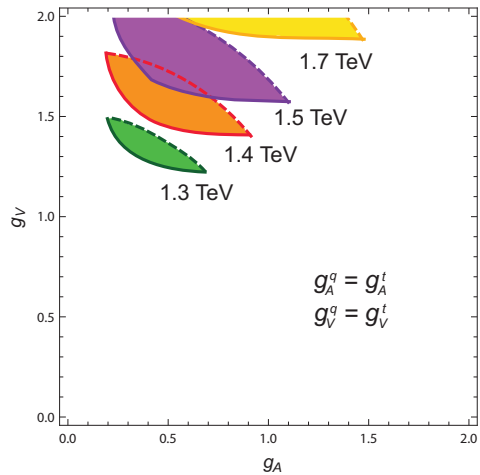}
\caption{\label{fig:universal} Contours at 95\% C.L. as 
a function of the vector and axial-vector couplings for different 
values of the resonance mass for flavor-universal couplings.}
\end{figure}

Our results are shown in Fig.~\ref{fig:universal}, where, for 
a given value of the mass of the color-octet we provide 
the allowed region at $95$\% C.L. in the $g_V-g_A$ plane. 
The solid lines are obtained from the charge asymmetry, 
while the dashed lines are derived from the last bin of the 
invariant mass distribution. 
The allowed regions are quite constrained; indeed at $90$\% C.L.,
we do not find any overlapping region for any value of the 
color-octet mass, and future experimental measurements with 
higher statistics can shrink significantly, or even exclude 
completely the allowed regions. With the most recent experimental 
values we find, in particular, that the asymmetric chiral color model
($g_V=\cot 2\theta, g_A=1/\sin 2\theta$, or $g_V=\sqrt{g_A^2-1}$) 
is disfavored. 

\begin{figure}[t]
\includegraphics[width=7cm]{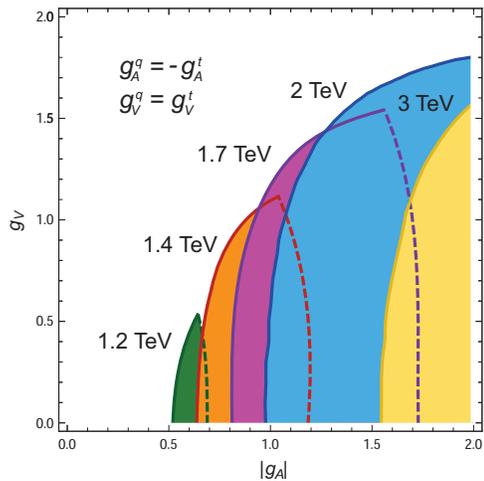}
\caption{\label{fig:alltogheter} Contours at 90\% C.L. as 
a function of the vector and axial-vector couplings for different 
values of the resonance mass and $g_A^q=-g_A^t$.}
\end{figure}

Another possibility to generate a positive charge asymmetry 
is to couple the third generation of quarks and the 
lighter quarks with axial-vector couplings of opposite sign:
$g_A^q=-g_A^t$. From \Eq{eq:bornqq} it is obvious that the 
actual sign of these couplings is irrelevant; only their 
relative sign is important because the asymmetric contributions 
to the differential cross section are proportional to their product.
Chiral color models with nonuniversal 
flavor couplings were already considered in the 
pioneering works~\cite{chiralcolor}. Our approach here
is, nevertheless, purely phenomenological, and building a realistic 
model in that scenario is beyond the scope of this paper. 
The results for the axigluon case with $g_A^q=-g_A^t=1$
are presented in Fig.~\ref{fig:anti}. That scenario 
is compatible with the experimental data for any mass
within $2\sigma$. The most general case is shown in 
Fig.~\ref{fig:alltogheter}. From Fig.~\ref{fig:alltogheter} 
and for $|g_A|<2$, we find that, independently of the 
resonance mass, the region about 
\beq
(|g_V|-2.3)^2+|g_A|^2 \gtrsim 1.8^2
\eeq
is excluded at 90\% C.L. Furthermore, 
for fixed values of the vector and axial-vector couplings 
the charge asymmetry sets a lower limit on the mass of the 
color-octet, while an upper bound can be set thanks to the 
invariant mass distribution, e.g. for $|g_A|=1$,
we find that at 90\% C.L.
\beq
1.33~{\rm TeV} < m_G < 2~{\rm TeV}~.
\eeq

\begin{figure}[t]
\includegraphics[width=7cm]{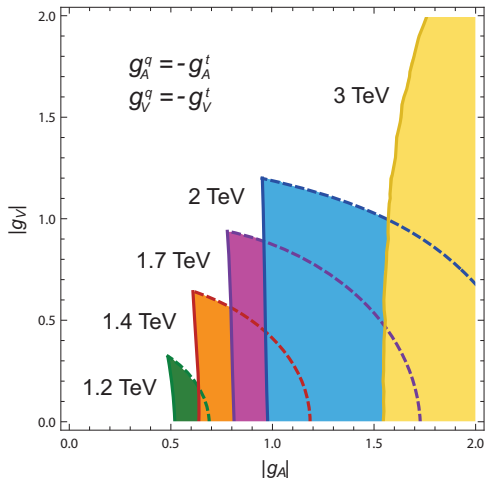}
\caption{\label{fig:antiflavor} Contours at 90\% C.L. as 
a function of the vector and axial-vector couplings for different 
values of the resonance mass and $g_V^q=-g_V^t$, $g_A^q=-g_A^t$.}
\end{figure}

Finally, we have also considered the case $g_V^q=-g_V^t$ and 
$g_A^q=-g_A^t$. Our results are presented in Fig.~\ref{fig:antiflavor}. 
Obviously, for $g_V=0$, we obtain the same result as in  
Fig.~\ref{fig:alltogheter}.

In conclusion, recent measurements of the charge asymmetry 
and the invariant mass distribution in top-antitop quark pair 
events allow for constraining the mass and couplings of 
hypothetic color-octet resonances decaying to top quarks 
with masses at the TeV scale. 
In the flavor-universal scenario, the allowed parameter 
space is quite constrained because the most recent measurements 
disfavor at $2\sigma$ vanishing or negative values of the 
charge asymmetry. In the flavor nonuniversal case,  
it is still possible to reconcile the experimental data 
with the existence of such resonances, and already a 
significant region of the parameter space can be 
excluded. In view of the significant progress over the last 
year from the experimental side, we expect that 
new results from Tevatron will further constrain 
efficiently the parameter space even before the 
start of the LHC, which is the natural place to discover 
those heavy resonances. At the LHC, there is no forward-backward 
asymmetry, obviously, but a sizeable charge asymmetry can be 
obtained by selecting events in the central 
region~\cite{Ferrario:2008wm}.

We thank the CERN TH TOP09 Institute, where this work was 
initiated, for hospitality and for many useful discussions. 
The work of P.F. is supported by the Consejo Superior de 
Investigaciones Cient\'ificas (CSIC). This work is also supported
by the Ministerio de Ciencia e Innovaci\'on under Grant
No. FPA2007-60323, by CPAN (Grant No. CSD2007-00042), 
by the Generalitat Valenciana under Grant No. PROMETEO/2008/069, 
and by the European Commission MRTN FLAVIAnet under Contract
No. MRTN-CT-2006-035482.

\end{document}